\documentclass[10pt,aps,twocolumn,showpacs,prd,preprintnumbers,floatfix,nofootinbib]{revtex4-1}

\usepackage{amsmath}
\usepackage{amsfonts}
\usepackage{graphicx}
\usepackage{amssymb,latexsym}
\usepackage{hyperref}
\usepackage{appendix}
\usepackage{multirow}

\begin{document}

\title{Low power on large scales in just enough inflation models}

\author{Erandy Ramirez}
\email{erandy@tum.de}

\affiliation{Excellence Cluster Universe, 
Technische Universit\"at M\"unchen, Boltzmannstr. 2, 85748 Garching, Germany.}

\begin{abstract}     
An early stage of kinetic energy domination for inflation is applied to  
single-field quadratic and hybrid-type potentials considering only the 
amount of inflation necessary to solve the problems of the standard
cosmological scenario. Using initial conditions inside the 1-sigma 
interval of the best-fit cosmological parameters for the potentials,
low values for the quadrupole component can be obtained.
\end{abstract}

\date{\today}

\preprint{}
\maketitle

\section{introduction}
Following previous works in which a modification of the chaotic inflation
scenario \cite{linde83} is proposed to quantify the predictions of 
the $\lambda\phi^4$ potential \cite{rs2}, 
in this work the idea is further applied to the $m^2\phi^2$ potential 
and a hybrid inflation potential during the stage when there is effectively 
a single field leading the dynamics. The intention is to evaluate what the 
effects of this scenario on different types of potentials are.
The implementation of this idea in both cases gives predictions for the 
inflationary parameters inside
the constraints imposed by observations \cite{K10}. As previously reported
in \cite{cpkl} where specific cutoffs of the primordial spectra are used,
an early stage of kinetic energy domination with only an amount of inflation
necessary to solve the problems of the standard cosmological scenario 
naturally gives rise to lower values of the first multipoles of the Cosmic
Microwave Background (CMB), this is also studied in \cite{vsb}. 
By doing a mode integration to find the best-fit 
values of this scenario for both potentials, initial conditions inside the 
1-sigma interval for the quantities that define this scenario of inflation 
can match the low value of the quadrupole although at a pivot scale two 
orders of magnitude bigger than $0.002\, {\rm Mpc}^{-1}$. Evaluation at scales 
closer to $0.05\, {\rm Mpc}^{-1}$ can also produce lower values of the 
quadrupole.

Additionally, an initial condition for the perturbations consistent with 
an early stage of kinetic energy domination is considered. 
The predictions and results for this case are in accordance to those 
found in \cite{ddr}, applied to the scenario already mentioned in \cite{cpkl},  
in which the form of the power spectrum does not show a significant
dependence on the choice of initial conditions. In the cases studied here, 
no previous stage of radiation domination is considered to set the initial 
conditions for the perturbations or the background evolution 
\cite{pk}, \cite{ws}. 

\section{notation and procedure}

This is a complementary work to \cite{rs2} and follows the same notation 
and procedure. This scenario considers the existence of an upper bound for 
the validity of a theory in terms of an effective potential, setting therefore
a limit to the total amount of inflation that can be produced. For the 
application of this scenario, a mode integration for the equations 
of inflationary perturbations has to be done
to find the behavior of the primordial power spectra during the initial
stage of kinetic energy domination. This early phase breaks slow-roll 
dynamics and a power-law parameterization for the spectra cannot be applied. 
From the mode integration one can use the resulting power spectra 
to obtain the values of the expansion in multipoles of the CMB anisotropies 
and the best-fit parameters for a specific potential.

The integration of the equations of motion for the background and 
perturbations is done with respect to the number of $e$-foldings $N$,
assuming homogeneity and isotropy from the beginning of inflation:  
\begingroup
\everymath{\scriptstyle}
\small
\begin{eqnarray}
\label{de}
\frac{dH}{dN} &=& \frac{V}{M_P^2 H}-3H \\ \nonumber
\frac{d\phi}{dN} &=& -\sqrt{6M_P^2-\frac{2V}{H^2}}.
\end{eqnarray}
\endgroup
$H$ is the Hubble rate of expansion, $V$ the potential energy and
$\phi$ the scalar field driving inflation. The conventions followed here are 
$\dot{\phi}<0\Rightarrow H'>0$ and $H \equiv dN/dt$, 
where $t$ is cosmic time, therefore $dN>0$ as $dt>0$. $M_P$ is the 
reduced Planck mass defined as $M_P\equiv\frac{m_{\rm pl}}{8\pi}$ and 
$m_{\rm pl} = 1.22\times 10^{19}$ GeV is the Planck mass. The notation 
is in units where $c=1, h =1$.

The equations of scalar and tensor perturbations are \cite{mfb}
\begingroup
\everymath{\scriptstyle}
\small
\begin{eqnarray}
\label{eN}
\frac{d^2 u_k^S}{d N^2}+(1-\epsilon_1)\frac{d u_k^S}{dN}
+\left[\left(\frac{k}{aH}\right)^2
-f_S(\epsilon_1,\epsilon_2,\epsilon_3)\right]u_k^S&=&0,
\\ \nonumber
\frac{d^2 u_k^T}{d N^2}+(1-\epsilon_1)\frac{d u_k^T}{dN}
+\left[\left(\frac{k}{aH}\right)^2-f_T(\epsilon_1)\right]u_k^T&=&0
\end{eqnarray}
\endgroup
where {\small $f_S(\epsilon_1,\epsilon_2,\epsilon_3)=2-\epsilon_1
+\frac{3}{2}\epsilon_2-\frac{1}{2}\epsilon_1\epsilon_2
+\frac{1}{2}\epsilon_2\epsilon_3+\frac{1}{4}\epsilon_2^2$} and
{\small $f_T(\epsilon_1)=2-\epsilon_1$}, in terms 
of the horizon flow functions defined as \cite{stg}
\begingroup
\everymath{\scriptstyle}
\small
\begin{eqnarray} 
\label{hff}
\epsilon_0\equiv\frac{H_i}{H},\quad \epsilon_{m+1}\equiv\frac{1}{\epsilon_m}
\frac{d\epsilon_m}{dN}, \quad m\ge 0. 
\end{eqnarray}
\endgroup
$u_k^{S,T}$ represents the usual
gauge-invariant combination of metric and field perturbations
for scalar and tensor fluctuations \cite{mfb}, $k$ is the comoving
wave number of each perturbation mode.

Following the procedure mentioned in \cite{rs2}, this scenario is applied 
here to two single-field inflation potentials :
\begingroup
\everymath{\scriptstyle}
\small
\begin{eqnarray} 
\label{potentials}
V = \frac{1}{2}m^2\phi^2,  \qquad
V = V_0\left(1+\frac{\phi^2}{16\pi M_P^2}\right)
\end{eqnarray}
\endgroup
where the second one corresponds to the hybrid inflation potential in the
notation of \cite{llms} for the stage where there is effectively one 
scalar field influencing the dynamics, assuming inflation finishes 
instantaneously. The conditions under which this happens are studied 
in \cite{sc}. In this case, as there is no natural end of inflation, the value 
$\phi=16.3M_P$ is used to stop the process of inflation.

To set the initial values for the perturbations, first Bunch--Davies vacuum 
initial conditions are considered:

scalars:
\begingroup
\everymath{\scriptstyle}
\small
\begin{eqnarray} 
\label{ic-bd-s}
u_k^S = \frac{1}{\sqrt{2k}}e^{-ik\tau_i}, 
\frac{du_k^S}{dN} = -i u_k^S\left(\frac{k}{a_iH_i}\right),
\end{eqnarray}
\endgroup

tensors:
\begingroup
\everymath{\scriptstyle}
\small
\begin{eqnarray} 
\label{ic-bd-t}
u_k^T = \frac{1}{\sqrt{k}}e^{-ik\tau_i}, \quad
\frac{du_k^T}{dN} = -i u_k^T\left(\frac{k}{a_iH_i}\right).
\end{eqnarray}
\endgroup

And an initial condition consistent with the solution of kinetic energy
domination for the mode equation is also considered \cite{cpkl}, \cite{ddr}:
\begingroup
\everymath{\scriptstyle}
\small
\begin{equation}
\label{ked}
u_k^{S,T}(\tau)=\sqrt{\frac{\pi}{8h_i}(1+2h_i\tau)}H_0^{(2)}\left(k\tau\right)
\end{equation}
\endgroup
where $\tau$ is conformal time defined as $d\tau=dt/a(t)$, $h=a,_{\tau}/a=aH$, 
$a$ the scale factor and $H_0^{(2)}$ is a Hankel function defined as 
$H_n^{(2)}(k\tau)\equiv J_n(k\tau)-iY_n(k\tau)$, with $J_n$ and $Y_n$ 
Bessel functions and $n$ a positive integer.
From this expression,
the initial conditions for the scalar and tensor perturbations considering
kinetic energy domination in terms of the number of $e$-foldings are

scalars:
\begingroup
\everymath{\scriptstyle}
\small
\begin{eqnarray} 
\label{ic-kd-s}
u_k^S &=& \sqrt{\frac{\pi}{8a_iH_i}}H_0^{(2)}\left(\frac{k}{2a_iH_i}\right), \\ 
\nonumber
\frac{du_k^S}{dN} &=& \sqrt{\frac{\pi}{8a_iH_i}}
\left[H_0^{(2)}\left(\frac{k}{2a_iH_i}\right)
-\frac{k}{a_iH_i}H_1^{(2)}\left(\frac{k}{2a_iH_i}\right)\right],
\end{eqnarray}
\endgroup

tensors:
\begingroup
\everymath{\scriptstyle}
\small
\begin{eqnarray} 
\label{ic-kd-t}
u_k^T &=& \sqrt{\frac{\pi}{4a_iH_i}}H_0^{(2)}\left(\frac{k}{2a_iH_i}\right), \\ 
\nonumber
\frac{du_k^T}{dN} &=& \sqrt{\frac{\pi}{4a_iH_i}}
\left[H_0^{(2)}\left(\frac{k}{2a_iH_i}\right)
-\frac{k}{a_iH_i}H_1^{(2)}\left(\frac{k}{2a_iH_i}\right)\right].
\end{eqnarray}
\endgroup

The expressions for the power spectra are \cite{llms}
\begingroup
\everymath{\scriptstyle}
\small
\begin{eqnarray}
\label{ps}
P_R = \frac{k^3}{4\pi^2M_p^2}\left|\frac{u_k^S}{z_S}\right|^2, \quad
P_T = \frac{2k^3}{\pi^2 M_p^2}\left|\frac{u_k^T}{z_T}\right|^2 
\end{eqnarray}
\endgroup

with $z_S = a\sqrt{\epsilon_1},\,\,\, z_T = a$. The scale factor is normalized
to 1 at the beginning of inflation.

In order to obtain the predictions of this scenario, 
the quantities used to set the initial conditions
for the background are the scalar field and the function $\epsilon_1$.
Since this scenario is a modification of chaotic inflation,  
the initial condition for the scalar field is assumed to be 
above the Planck scale,
at a value that assures to produce only the amount of inflation necessary to
solve the problems of the standard cosmological scenario. The initial value
of the function $\epsilon_1$ sets the phase of kinetic energy domination
from which the system of equations rapidly joins the slow-roll attractor
regime. Consequently, the potential is many orders of magnitude smaller
than $M_P^4$ when inflation starts.

The best-fit parameters are obtained by reusing the existing parameters 
in the publicly available CAMB \cite{lcl} and COSMOMC codes 
\cite{lb}. Once the initial value for the field, $\epsilon_1$ and the
coefficients of the potentials are set, the start of inflation determines
the initial scale $k$ that will be used
for the mode integration and extrapolation of the primordial spectra for the
CAMB code assuming sudden reheating to transfer the scales from GeV to
$Mpc^{-1}$. For each potential a new covariance matrix is found.  

When the coefficients $m$ and $V_0$ in the potentials are varied, 
this scenario contains one parameter extra than a standard $\Lambda$ 
cold dark matter scenario ($\Lambda$CDM) with 6 primary parameters. 
For these cases, the Akaike information criterion and a $\chi^2$ 
criterion are applied in order to know the goodness of fit for this scenario.

\section{results}

The best-fit parameters are calculated for each potential in three cases
corresponding to Bunch-Davies initial conditions
for the perturbations and kinetic energy domination, varying the 
coefficients $m$ and $V_0$ with tensors included and keeping them
fixed without tensor modes. To compare with the results found 
in \cite{rs2}, a simulation for the $\lambda\phi^4$
potential was also performed when the solution  for kinetic
energy domination is applied to set the initial conditions for the
perturbations.

The value of the pivot scale in the Monte Carlo code does not affect 
the mode integration since the moment when inflation starts with 
kinetic energy domination determines the initial scale for the integration. 
Nonetheless, a simulation was done for each potential in order 
to check that the best-fit parameters were indeed not affected
by changing the value of the pivot scale in the COSMOMC code.

In Table~\ref{results} the results of the Monte Carlo simulations
are shown for the potentials in Eq.~(\ref{potentials}); all simulations
have been done for 6 chains of 100 000 samples of length each, using
WMAP 7-year data including TT, TE and EE spectra only. To account for the
number of final independent samples, $20\%$ of rows are excluded
from the analysis of the chains. The lensing effect is taken into
account in all cases. The values in Table~\ref{results} for the 
coefficients of the potentials are $m/M_p$ and $V_0/M_P^4$ respectively.

\begin{table*}
\caption{Results of Monte Carlo integration for kinetic energy domination 
(k. e. d.) and Bunch--Davies (BD) vacuum initial conditions. The sixth column
 shows the values for coefficients of the potentials.}
\begin{tabular}{|c|c|c|c|c|c|c|c|c|c|}\hline
\label{results}
 &
 &
Initial &
 &
 &
 &
 &
independent &
 &
 \\ 

  &
Potential &
conditions &
$\epsilon_{1,i}$ &
$\phi_i/M_p$ &
Coefficient &
Tensors &
samples &
Burn-in &
R-1\\ \hline

 1 &
$\frac{1}{2}m^2\phi^2$ &
k. e. d. &
$[2.63,2.999]$ &
$[16.57,17.97]$ &
$[6.14\times 10^{-6},7.13\times 10^{-6}]$ &
included &
6985 &
220 &
0.0017 \\ \hline

 2 &
  &
BD &
same &
same &
same &
included &
5749 &
249  &
0.0028 \\ \hline

 3 &
  &
BD &
same &
same &
$6.62\times 10^{-6}$ &
not included &
5828 &
164  &
0.0021 \\ \hline

 4 &
hybrid &
k. e. d. &
$[2.9,2.999]$ &
$[23.6,24.4]$ &
$[1.73\times 10^{-10},2.12\times 10^{-10}]$ &
included &
5711 &
324  &
0.0022 \\ \hline

 5 &
 &
BD &
same &
same &
same &
included &
3359 &
212 &
0.0048 \\ \hline
\end{tabular}
\end{table*}

According to the conclusions of \cite{rs2}, the most favored situation
for this scenario of inflation corresponds to $\epsilon_{1,i}$ being as 
close as possible to 3. The results in Table~\ref{results} show the biggest
number of final independent samples when the coefficients of both potentials
are varied, the tensor modes are included and an initial condition with
kinetic energy domination for the perturbations is used. In the case 
of the hybrid potential, the prior interval 
of initial values for $\epsilon_1$ that allowed to conclude successfully the 
simulations had to be reduced to $[2.9,2.999]$, whereas for
the $\phi^2$ potential  a wider range of initial values 
for this function is allowed. For all cases a simulation with a pivot scale 
equal to $0.01\, {\rm Mpc}^{-1}$ is performed in order to check that 
the best-fit
values are not too different from those with 
$k_{\rm pivot} = 0.05\,{\rm Mpc}^{-1}$ 
from the same prior intervals. Only the final number of independent samples 
depends significantly on the value of the pivot scale. Its value almost 
doubles for $k_{\rm pivot} = 0.01\,{\rm Mpc}^{-1}$ with respect 
to $0.05\,{\rm Mpc}^{-1}$. 

\begin{table*}
\caption{Best-fit cosmological parameters for some of the examples 
in Table~\ref{results}. The parameter $C_P$ represents 
the coefficient of the potential divided by $M_p$ and $M_P^4$ for each
potential respectively.}
\begin{tabular}{|c|c|c|c|c|c|c|c|c|c|c|c|c|c|c|c|}\hline
\label{results2}
  &
  &
  &
 $\Omega_b h^2$ &
 $\Omega_{DM}$ &
 $\theta$ &
 $\tau$ &
 $\phi_i/M_p$ &
 $\epsilon_{1,i}$ &
 $\ln(10^{10} C_P)$ &
 $\Omega_{\Lambda}$  &
 Age/Gyr &
 $\Omega_m$ &
 $z_{re}$ &
 $r_{10}$ &
 $H_0$  \\ \hline

$\frac{1}{2}m^2\phi^2$  &
1,2 &
$\mu$  &
0.022  &
0.11  &
1.039  &
0.087  &
17.47  &
2.78  &
11.1  &
0.73  &
13.8 &
0.27 &
10.48 &
0.067 &
70.52 \\ \hline

 &
 &
$\sigma$  &
0.0003  &
0.005  &
0.002  &
0.01  &
0.3  &
0.09  &
0.02  &
0.03  &
0.08 &
0.03 &
1.2 &
0.002 &
1.9 \\ \hline

  &
3 &
$\mu$ &
0.023  &
0.11  &
1.039  &
0.09  &
17.47  &
2.78  &
11.1 &
0.72  &
13.8  &
0.28  &
11.0  &
-  &
69.9   \\ \hline

 &
 &
$\sigma$  &
0.0003  &
0.005  &
0.002  &
0.01  &
0.3  &
0.09  &
- &
0.02  &
0.08 &
0.02 &
0.7 &
- &
1.6 \\ \hline

hybrid &
4,5 &
$\mu$  &
0.023  &
0.11  &
1.041  &
0.1  &
24.1  &
2.93  &
0.63  &
0.75  &
13.6  &
0.25  &
11.04  &
0.03  &
73.13 \\ \hline

 &
 &
$\sigma$  &
0.0003  &
0.005  &
0.002  &
0.02  &
0.2  &
0.02  &
0.03  &
0.02  &
0.1 &
0.02 &
1.2 &
0.001 &
1.9 \\ \hline
\end{tabular}
\end{table*}

In Table~\ref{results2}, the best-fit cosmological parameters  are
presented when the coefficients of both potentials are varied and 
tensor modes are included. An initial condition with kinetic energy domination
for the perturbations and for Bunch--Davies vacuum shows no  
differences in the predictions for the parameters. In the case of the
hybrid potential, only the results for kinetic energy domination are
presented since there are no changes in those values as compared to
using Bunch-Davies initial conditions. The same situation applies to
the results of the $\lambda\phi^4$ potential
when compared to the values for the cosmological parameters found 
in \cite{rs2}, and therefore the results are omitted here.
The distributions of the cosmological parameters for case 1 in 
Table~\ref{results} are presented in the Appendix.
It can be observed that, as in \cite{rs2}, the initial value of the field 
and $\epsilon_{1,i}$ are degenerate with each other and the only independent 
parameter for this scenario is the total amount of inflation produced by 
each model. These two quantities have non Gaussian distributions whereas 
the distribution for the coefficient of the potential is Gaussian. 

As mentioned before, all models with the exception of case 3 presented
in Table~\ref{results}, are described by 7 primary parameters, one more
than a standard $\Lambda$CDM model with 6 parameters. For these
cases the Akaike information criterion (AIC) is applied in order 
to compare the goodness of fit: ${\rm AIC} = -2{\cal L}_{\rm max}
+2N_{\rm par}$ where ${\cal L}_{\rm max}$ is the maximum likelihood of 
each model and $N_{\rm par}$ is the number of parameters. 
The results are shown in Table~\ref{results_1}. The values of 
$\Delta {\rm AIC}$ 
for both potentials here are smaller than those for 
the $\lambda\phi^4$ potential as seen in \cite{rs2}, the difference 
for the $m^2\phi^2$ potential is the smallest of the three 
cases. Therefore, the results for these two potentials in this scenario 
are only marginally worse than the standard $\Lambda$CDM scenario.

\begin{table}
\caption{Akaike information criterion applied to the models presented in
Tables~\ref{results},\ref{results2} with respect to a $\Lambda$CDM model
of 6 parameters.}
\begin{tabular}{|c|c|c|c|c|c|}\hline
\label{results_1}
Model &
$-\ln {\cal L}_{\rm max}$ &
$N_{\rm par}$ &
AIC &
$\Delta$AIC &
$\chi^2$ \\ \hline

$\Lambda$CDM &
3737.2 &
6 &
7486.4 &
0 &
7474.4 \\ \hline

1 &
3736.384 &
7 &
7486.768 &
0.368 &
7472.768 \\ \hline

2 &
3736.39 &
7 &
7486.78  &
0.38  &
7472.78 \\ \hline

4 &
3737.419  &
7 &
7488.838 &
2.07  &
7474.838 \\ \hline

5 &
3737.509 &
7 &
7489.018  &
2.25  &
7475.018 \\ \hline
\end{tabular}
\end{table}

\subsection{Mode Integration, Spectral index and Running}
The results of the mode integration are presented
only for the $m^2\phi^2$ potential considering two possibilities : 
Equations~(\ref{ic-bd-s}), (\ref{ic-bd-t}), (\ref{ic-kd-s}) and 
(\ref{ic-kd-t})  for the initial conditions of the perturbations. 
The behavior of the spectral indices $n_{S,T}$, and runnings 
$\alpha_{S,T}$ for scalar and tensor perturbations using the mode integration 
is also shown. Two expressions 
that can be applied when the power spectra are continuous functions 
are written in the appendix. When they are
obtained from a mode integration, they are discrete functions as a
result of the integration of Eqs.~(\ref{de}) and (\ref{eN}) for 200 
fixed values of the scale $k$ and an alternative way to evaluate 
the indices and runnings is used. The scales for the mode
integration were determined using the first mode, obtained
when inflation starts, up to a value five orders of magnitude bigger,
spacing the interval into 200 modes.

\begin{figure*}
\begin{centering}
\includegraphics[height=2.4in,width=3.4in,angle=0]{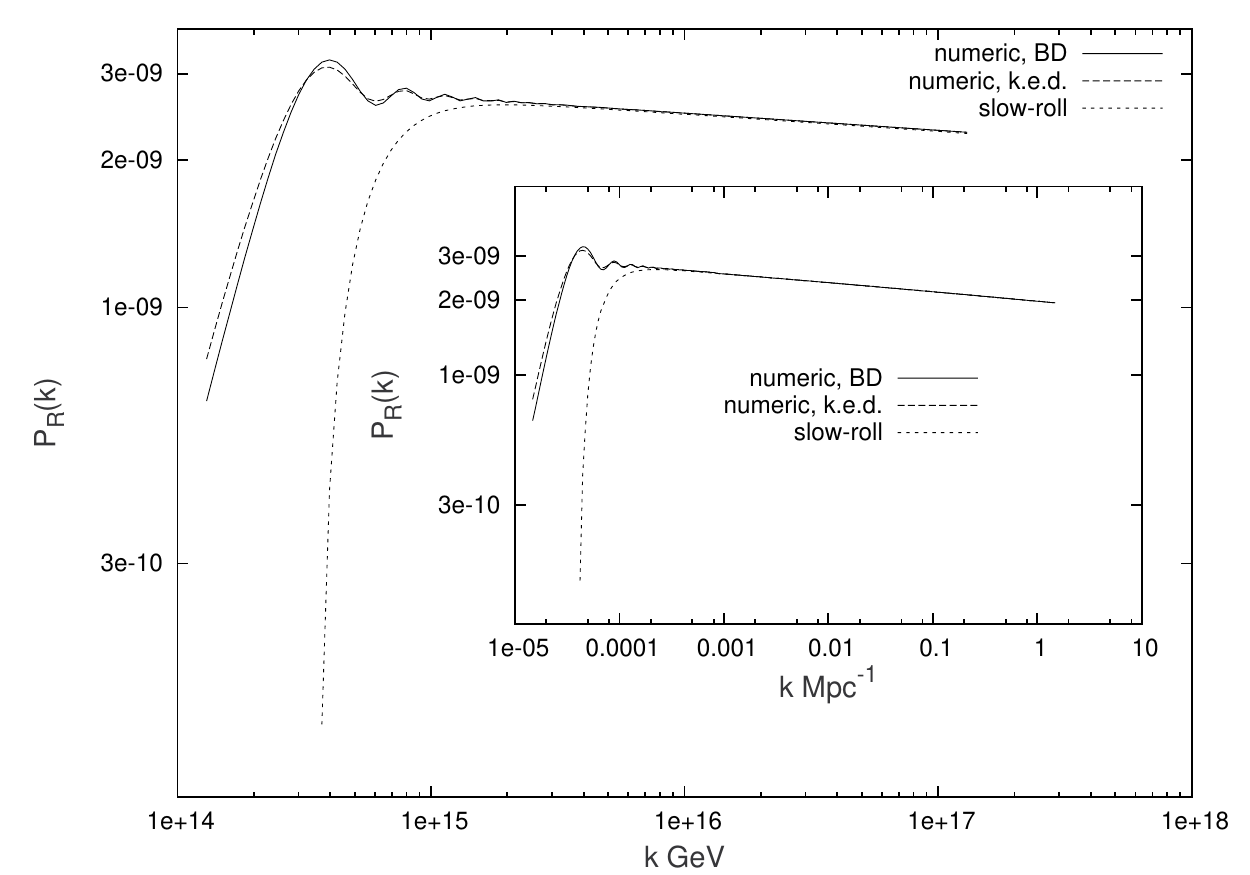}
\includegraphics[height=2.4in,width=3.4in,angle=0]{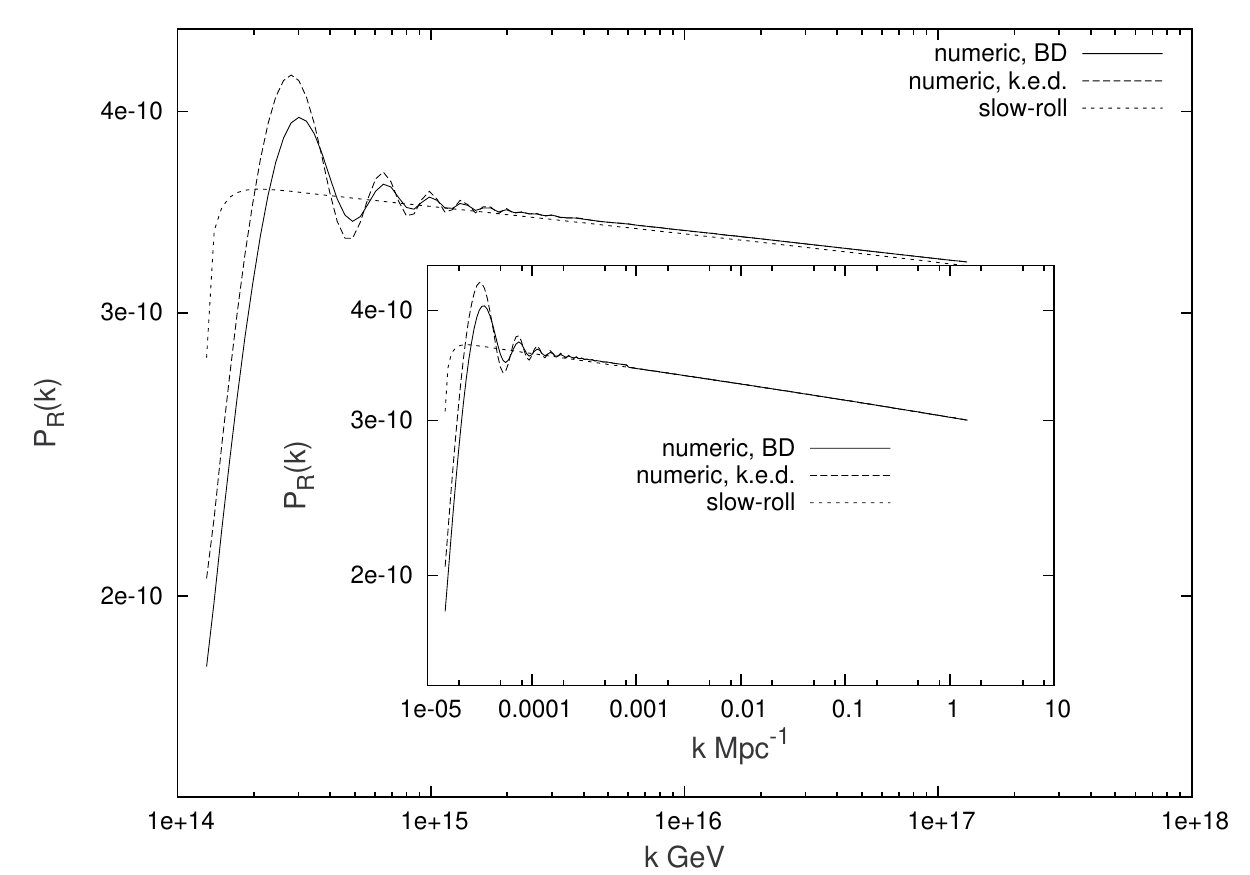}
\caption{Mode integration for the scalar and tensor modes from 
the best-fit value $\epsilon_{1,i} =2.78$, $\phi_i/M_p = 17.47$, 
$6.55\times 10^{-6}$ with tensors, cases 1 and 2 in Table~\ref{results}.}
\label{mode1_s}
\end{centering}
\end{figure*}

The results of the mode integration for both power spectra are shown 
in Fig.~\ref{mode1_s}, they show their behavior for 
Bunch--Davies vacuum initial conditions, for kinetic energy domination
and for slow-roll at first-order. 
Both initial conditions do not differ significantly from each other when
inflation starts. Therefore, as the process of inflation goes on, deeper 
sub horizon modes coincide with the attractor slow-roll behavior 
when they are frozen outside the horizon without 
being affected by different initial conditions even having approximately only  
60 $e$-folds of total accelerated expansion. The oscillations 
in the numeric solutions appear because, as mentioned in \cite{rs2}, the first 
modes being integrated are already nearly at the moment of horizon crossing 
and the oscillatory solutions of the mode equation can be observed. 

The behavior of the spectral index and its running 
for the scalar modes with the expressions written in the appendix are 
shown in Fig.~\ref{mode2_s}. In the case of the spectral
index, the plot shows the result of the mode integration from the second
mode onwards and for the running from the third mode onwards. This is 
only due to the expression taken to approximate the 
derivatives of the discrete power spectrum. However, both initial 
conditions seem to suggest that the initial value of the spectral index 
is close to or at 3, from which the oscillations of the solutions of the
mode equation are also present in the derivatives. A similar behavior is 
observed for the running. In the case of the spectral index, the derivative 
is always positive.
The behavior of the same quantities for the tensor modes are shown for
completeness in the Appendix.

\begin{figure*}
\begin{centering}
\includegraphics[height=2.4in,width=3.4in,angle=0]{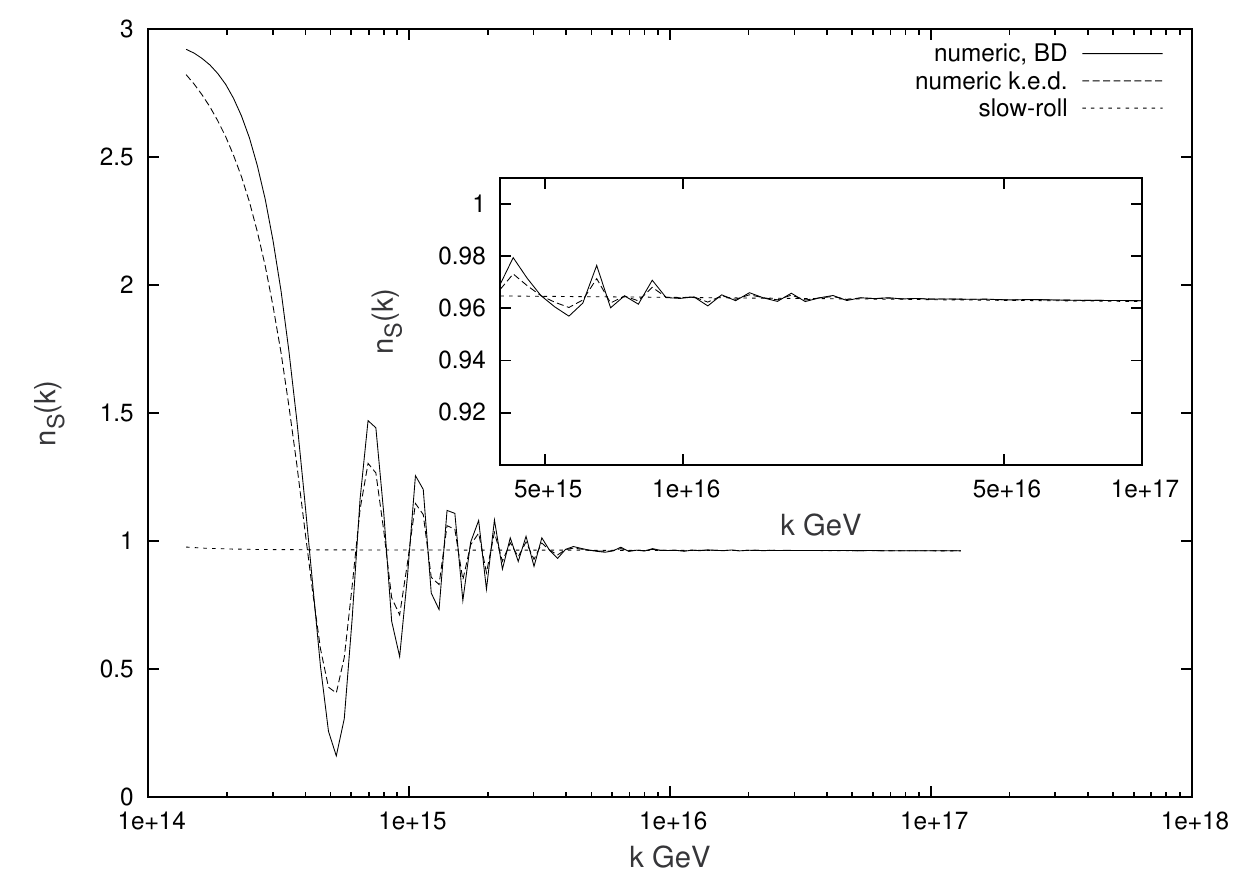}
\includegraphics[height=2.4in,width=3.4in,angle=0]{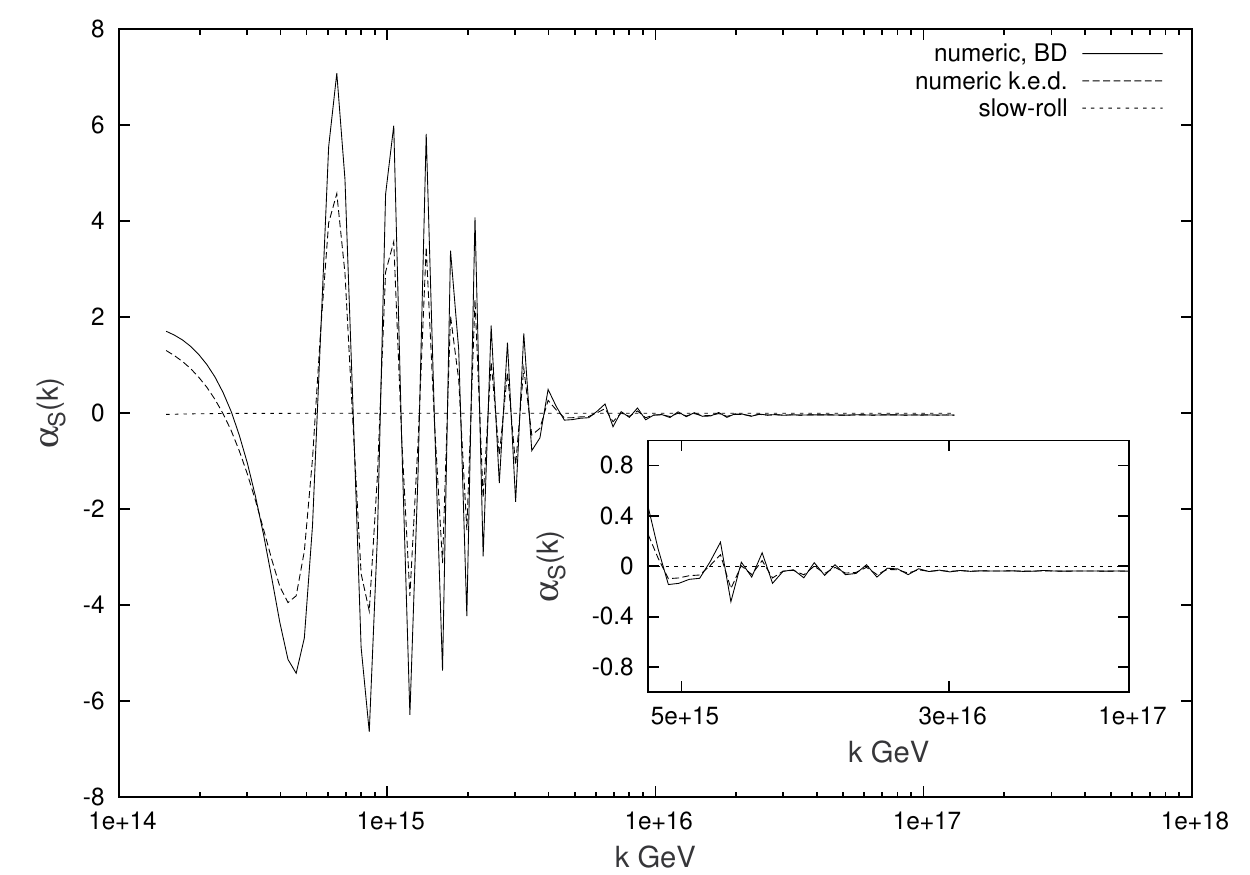}
\caption{Behavior of the spectral index and its running for the scalar modes 
using the mode integration, from the best-fit value 
$\epsilon_{1,i} =2.78$, $\phi_i/M_p = 17.47$,  $6.55\times 10^{-6}$ 
with tensors, showing cases 1 and 2 in Table~\ref{results}, compared to 
the slow-roll prediction at first order for the same initial condition.}
\label{mode2_s}
\end{centering}
\end{figure*}

To quantify the predictions of each potential for the best-fit values 
obtained in the Monte Carlo integration, the slow-roll expressions for the 
amplitude of the scalar spectrum, the spectral index and the running at 
first-order in slow-roll are used. They are presented in Table~\ref{results3}
for two pivot scales at 0.01,0.05 $Mpc^{-1}$, the first three cases
of Table~\ref{results}, \ref{results2} give the same predictions since the
best-fit values for the initial $\epsilon_1$ and $\phi/M_P$
are the same. They satisfy the constraint on the amplitude of scalar 
perturbations at 1 and 2 sigma levels for the analysis of WMAP7 data
including running of the spectral index.

\begin{table*}
\caption{Values for inflationary observables at two different pivot scales
for the best-fit values presented in Table~\ref{results2}, $r$ is the 
tensor-to-scalar ratio.}
\begin{tabular}{|c|c|c|c|c|c|c|c|}\hline
\label{results3}
  &
Best-fit values: $\epsilon_{1,i}$, $\phi_i/M_P$, $C_P$ &
$N_T$ &
$k_*(Mpc^{-1}), N_*$ &
 r &
$n_s$ &
$dn_s/dlnk$ &
Amplitude \\ \hline

1,2,3 &
2.78, 17.47, $6.55\times 10^{-6} M_P$ &
63.2 &
0.05,  58.3 &
0.14 &
0.97 &
$-6.8\times 10^{-4}$ &
$2.45\times 10^{-9}$ \\ \hline

 &
 &
 &
 0.01,  59.8 &
0.13 &
0.97 &
$-8.99\times 10^{-3}$ &
$2.57\times 10^{-9}$ \\ \hline

4,5 &
2.93, 24.1, $1.88\times 10^{-10} M_P^4$ &
61.85 &
0.05, 56.8 &
0.056 &
0.97 &
$-1.7\times 10^{-4}$ &
$2.33\times 10^{-9}$ \\ \hline

 &
 &
 &
0.01, 58.2 &
0.055 &
0.99 &
$-6.1\times 10^{-3}$ &
$2.37\times 10^{-9}$\\ \hline
\end{tabular}
\end{table*}

\subsection{Low Multipoles}

An early epoch of kinetic energy domination has been proposed as 
an explanation for the smaller than expected power of the low multipoles 
of the CMB anisotropies \cite{rs2}, \cite{cpkl}, \cite{vsb}, \cite{ddr}, 
\cite{pk}, \cite{ws}, \cite{chss}, \cite{sshc}, \cite{hs}. By applying initial 
conditions for the field, $\epsilon_{1,i}$ and the coefficient of the 
potential inside the 1$\sigma$ interval of the best-fit values in 
Table~\ref{results2}, it is possible to find the quadrupole below 200 
with values of the tensor-to-scalar ratio, spectral index, running and 
amplitude of scalar perturbations inside the 1 and 2$\sigma$ intervals allowed 
by observations for the analysis with tensors and running \cite{K10}. However, 
this happens evaluating these quantities at a scale of $\sim 0.3$ 
${\rm Mpc}^{-1}$. 
As the initial conditions give higher values of the quadrupole, 
the inflationary predictions are met for values of the scale closer 
to 0.5 ${\rm Mpc}^{-1}$. Some examples of this for each of the potentials 
in Eq.~(\ref{potentials}) are provided in Table~\ref{results4}. The first 
and fifth lines show the predictions for the quadrupole using the central 
values of the best-fit parameters. As for the other examples, one can 
observe that small changes in the field and $\epsilon_{1,i}$ have a more 
significant effect than changing the value of the coefficient of the 
potential. The same is applicable to the results of the hybrid potential.

\begin{table*}
\caption{Values for the quadrupole and octopole for the best-fit values, 
of 1,2, 4 and 5 in Table~\ref{results2}.}
\begin{tabular}{|c|c c c c c c c|c c c c c c|}\hline
\label{results4}
 &
 & 
$\phi_i/M_P$ &
$\epsilon_{1,i}$ &
$C_P$ &
$N_T$ & 
Quadrupole &
Octopole  &
$r$ &
$n_s$ &
$dn_s/d\ln k$ &
Amplitude &
$N_* $ &
$ k_*({\rm Mpc}^{-1})$ \\ \hline\hline
 
\multirow{4}{*}{$m^2\phi^2$} &
  1&
17.47 &
2.78 &
$6.55\times 10^{-6} M_P$ &
63.2 &
1120.6 &
1050.0 &
0.14   &
0.97 &
$-6.8\times 10^{-4}$  &
$2.45\times 10^{-9}$   &
 58.3 &
 0.05 \\ 

 &
 2  &
16.94 &
2.78 &
$6.57\times 10^{-6}M_P$ &
59.1 &
194.94 &
209.04  &
0.14 &
0.99 &
-0.066 & 
$2.27\times 10^{-9}$  &
56.35  &
0.35  \\ 

 &
 3  &
17.3 &
2.85 &
$6.56\times 10^{-6}M_P$ &
60.52 &
656.62 &
717.42  &
0.14 &
0.99 &
-0.08 & 
$2.36\times 10^{-9}$  &
57.9  &
0.08  \\ 

 &
 4  &
17.35 &
2.85 &
$6.56\times 10^{-6}M_P$ &
60.9 &
820.91 &
919.51  &
0.14 &
0.99 &
-0.06 & 
$2.4\times 10^{-9}$  &
58.2  &
0.057  \\ \hline \hline

\multirow{4}{*}{hybrid} &

5 &
24.1 &
2.93 &
$1.88\times 10^{-10} M_P^4$ &
61.85 &
1035.9 &
973.25  &
0.056 &
0.99 &
$-2\times 10^{-4}$ &
$2.33\times 10^{-9}$ &
56.79 &
0.053 \\

 &
6 &
23.9 &
2.95 &
$1.88\times 10^{-10} M_P^4$ &
57.71 &
213.42 &
220.11  &
0.057 &
1.0 &
-0.041 &
$2.23\times 10^{-9}$ &
54.7 &
0.42 \\ 

 &
7 &
24.0 &
2.94 &
$1.88\times 10^{-10} M_P^4$ &
59.86 &
942.37 &
995.96  &
0.056 &
1.0 &
-0.037 &
$2.3\times 10^{-9}$ &
56.8 &
0.052 \\ 

 &
8 &
23.95 &
2.94 &
$1.83\times 10^{-10} M_P^4$ &
59.25 &
683.31 &
753.49  &
0.056 &
0.99 &
-0.037 &
$2.22\times 10^{-9}$ &
56.2 &
0.094 \\ \hline
\end{tabular}
\end{table*}

\begin{figure}
\begin{centering}
\includegraphics[height=2.4in,width=3.4in,angle=0]{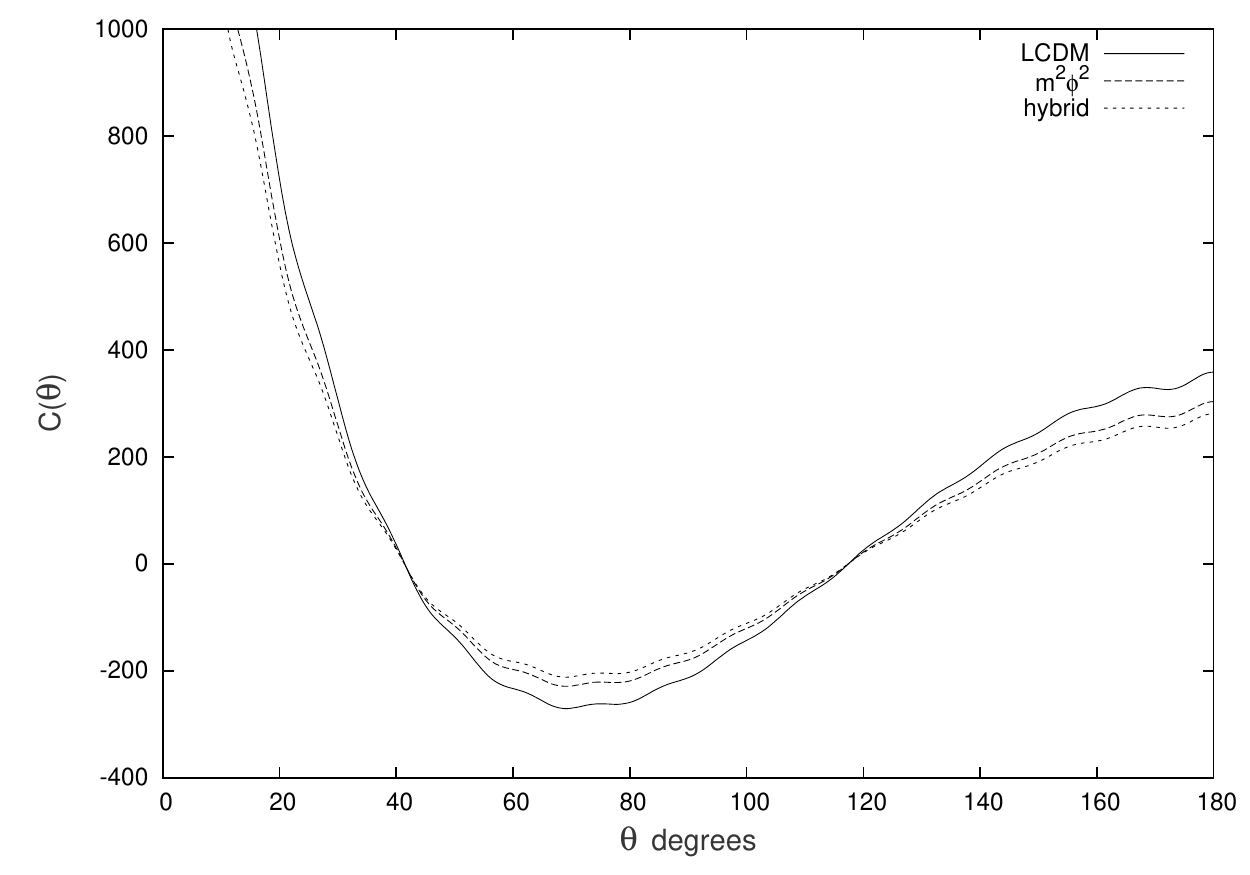}
\caption{Two point angular correlation function for the best-fit values 
of cases 1,2, 4 and 5 in Table~\ref{results2}, and a $\Lambda$CDM model
with $n_s = 0.96$.}
\label{pcf}
\end{centering}
\end{figure}

Although the value for the quadrupole predicted by the best-fit models
is  not low enough on scales relevant for observations, the suppression 
of power has more statistical significance for the two-point angular 
correlation function than for the 
quadrupole \footnote{Many thanks to Dominik Schwarz for pointing this out.} 
\cite{sshc}, \cite{shcss}. The two-point correlation function is plotted 
in Fig.~\ref{pcf} for the central values in Table~\ref{results2}, the
expression used is:
\begingroup
\everymath{\scriptstyle}
\small
\begin{eqnarray}
\label{2pcf}
C(\theta)=\frac{1}{4\pi}\sum_{\ell =2}^{\ell_{\rm max}} 
\left(2\ell+1\right)C_{\ell}P_{\ell}\left({\rm cos}\,\theta\right).
\end{eqnarray}
\endgroup
where $C_{\ell}$ is the angular power spectra and
$P_{\ell}$ Legendre polynomials. The expansion is done for 
$\ell_{\rm max} = 30$. For angles smaller than approximately $40^{\circ}$, 
the predictions for this function with the central values of the 
best-fit models are in general more suppressed than those of the 
$\Lambda$CDM model. 

\section{discussion}
The implementation of the $m^2\phi^2$ and hybrid potentials for an early 
stage of kinetic  energy domination with only 
the sufficient amount of accelerated expansion gives results consistent 
with those found before for the $\lambda\phi^4$ potential. The distributions 
of the quantities that set the initial conditions for this
scenario, that is the initial values of the scalar field
and the function $\epsilon_1$, show a degeneracy between each other, which
confirms that the only independent parameter for this scenario is the 
total amount of inflation produced.
The mode integration for these potentials shows again the presence of a 
cutoff of power on large scales. The spectral indices and the runnings
obtained from the mode integration show the oscillations of the  solutions of
the mode equations of perturbations before the system joins the inflationary
slow-roll attractor.
In accordance with the results of \cite{rs2}, initial
conditions inside the 1$\sigma$ interval of the best-fit values for the 
field and $\epsilon_1$ can 
give a significant suppression of power on the largest scales, and reproduce
the value of the quadrupole at very large scales. At  
$k_*=0.05\, {\rm Mpc}^{-1}$ 
there is already a suppression of power but not large enough to be consistent 
with observations. Additionally, the behavior of the two-point angular 
correlation function shows that on large angular scales, the power is 
suppressed for the best-fit models of the $m^2\phi^2$ and hybrid potentials 
with respect to the $\Lambda$CDM case. 

As mentioned before, the pivot scale set to do the Monte Carlo
integration must not alter the results for the best-fit parameters as was found
here by running a simulation for the same intervals with a different value
of this quantity ($0.01 {\rm Mpc}^{-1}$). What did change was the final number 
of independent samples, which in some cases is almost twice the amount 
of them as compared to a pivot scale of $0.05\, {\rm Mpc}^{-1}$. Whether this 
has a physical significance is not further investigated here.  

Another aspect of this work is the use of initial conditions for the 
perturbations consistent with an early stage of kinetic energy domination. 
The results of the mode integration show no significant alteration with 
respect to the usual initial conditions in the vacuum, the solutions 
rapidly join the slow-roll attractor and they have in fact a less sharp 
cutoff of power on large scales. This however, does not affect the values of
the best-fit parameters obtained from the same prior intervals for 
Bunch--Davies vacuum initial conditions and a homogeneous initial condition 
in kinetic energy domination. 

Another point to mention is, that similarly to the results for the 
$\lambda\phi^4$ potential, the biggest number of final independent samples
was found for those initial conditions with $\epsilon_{1,i}\simeq 3$. This
perhaps can again be interpreted as an indication that this initial
condition represents the most favored situation for this scenario. 

Since the use of initial conditions for the perturbations different from
the vacuum does not alter the predictions of this scenario, an interesting
possibility for further exploration would be its application to multi field
inflation scenarios. For these cases, perhaps the effect of the suppression 
of power on large scales could have a more significant influence on the 
predictions of this scenario.

\section{acknowledgments}
This research was supported by the DFG cluster of excellence 
"Origin and Structure of the Universe''. It is a pleasure to thank 
Dominik Schwarz, Houri Ziaeepour  and Christoph R\"ath for comments 
and revision of the manuscript. I thank Sandipan Kundu for interesting 
comments and discussion and H. J. de Vega and S\'ebastien Clesse for hints to 
the literature. I acknowledge the use of the Linux Cluster of 
the Leibniz-Rechenzentrum der Bayerischen Akademie der Wissenschaften.
The use of the CAMB and COSMOMC packages as well as WMAP 7-yr data 
from the LAMBDA server are also acknowledged.

\appendix
\numberwithin{equation}{section}

\section{Spectral indices and runnings}

Applying the definition of the spectral indices and their runnings, 
as given for example in \cite{llms}:
\begingroup
\everymath{\scriptstyle}
\small
\begin{eqnarray} 
\label{ir-def}
n_S(k)-1\equiv\frac{d \ln P_R}{d \ln k}, \quad 
n_T\equiv(k)\frac{d \ln P_T}{d \ln k}, \\ \nonumber
\alpha_S(k)\equiv\frac{d n_S}{d \ln k}, \quad
\alpha_T(k)\equiv\frac{d n_T}{d \ln k},
\end{eqnarray}
\endgroup
where $P_S$ and $P_T$ are given by Eq.~(\ref{ps}), one arrives to the
following expressions for the spectral indices:
\begingroup
\everymath{\scriptstyle}
\small
\begin{eqnarray} 
\label{index-exp}
n_S -1 &=& \frac{1}{(1-\epsilon_1)}\left[1-3\epsilon_1-\epsilon_2
+2\frac{d\ln|u_k^S|}{dN}\right] \\ \nonumber
n_T &=& \frac{1}{(1-\epsilon_1)}\left[1-3\epsilon_1
+2\frac{d\ln|u_k^T|}{dN}\right] 
\end{eqnarray}
\endgroup
and for the runnings :
\begingroup
\everymath{\scriptstyle}
\small
\begin{eqnarray} 
\label{runnings-exp}
\alpha_S &=&\frac{1}{(1-\epsilon_1)^2}\left[2\frac{d^2\ln|u_k^S|}{dN^2}
-\epsilon_2(3\epsilon_1+\epsilon_3) \right. \nonumber \\ 
&& \qquad \qquad \qquad \qquad \qquad
\left. +\epsilon_1\epsilon_2(n_s-1)\right]\\ \nonumber 
\alpha_T &=&\frac{1}{(1-\epsilon_1)^2}\left[2\frac{d^2\ln|u_k^T|}{dN^2}
+\epsilon_1\epsilon_2(n_T-3)\right]
\end{eqnarray}
\endgroup
evaluated at $k=aH$, the moment of horizon crossing for each 
perturbation mode. If one substitutes the
expressions of $u_k^{S,T}$ using the Stewart--Lyth solutions for the mode
equations \cite{sl}:
\begingroup
\everymath{\scriptstyle}
\small
\begin{eqnarray} 
\label{sols}
|u_k^S|&\simeq& \frac{1}{\sqrt{aH}}\left[1-\epsilon_1(1+C)
-\frac{C}{2}\epsilon_2\right], \\ \nonumber
|u_k^T|&\simeq&\frac{1}{\sqrt{2aH}}\left[1-(1+C)\epsilon_1\right],
\end{eqnarray}
\endgroup
where $C=-2+\ln 2+\gamma\simeq-0.73$, one recovers the usual 
slow-roll expressions at first-order:
\begingroup
\everymath{\scriptstyle}
\small
\begin{eqnarray} 
\label{sr-e}
n_S\simeq 1-2\epsilon_1-\epsilon_2, \quad 
\alpha_S\simeq -\epsilon_2(2\epsilon_1+\epsilon_3),\\ \nonumber
n_T\simeq-2\epsilon_1, \qquad \qquad \qquad \quad 
\alpha_T\simeq-2\epsilon_1\epsilon_2.
\end{eqnarray}
\endgroup

\begin{figure*}
\begin{centering}
\includegraphics[height=2.4in,width=3.2in,angle=0]{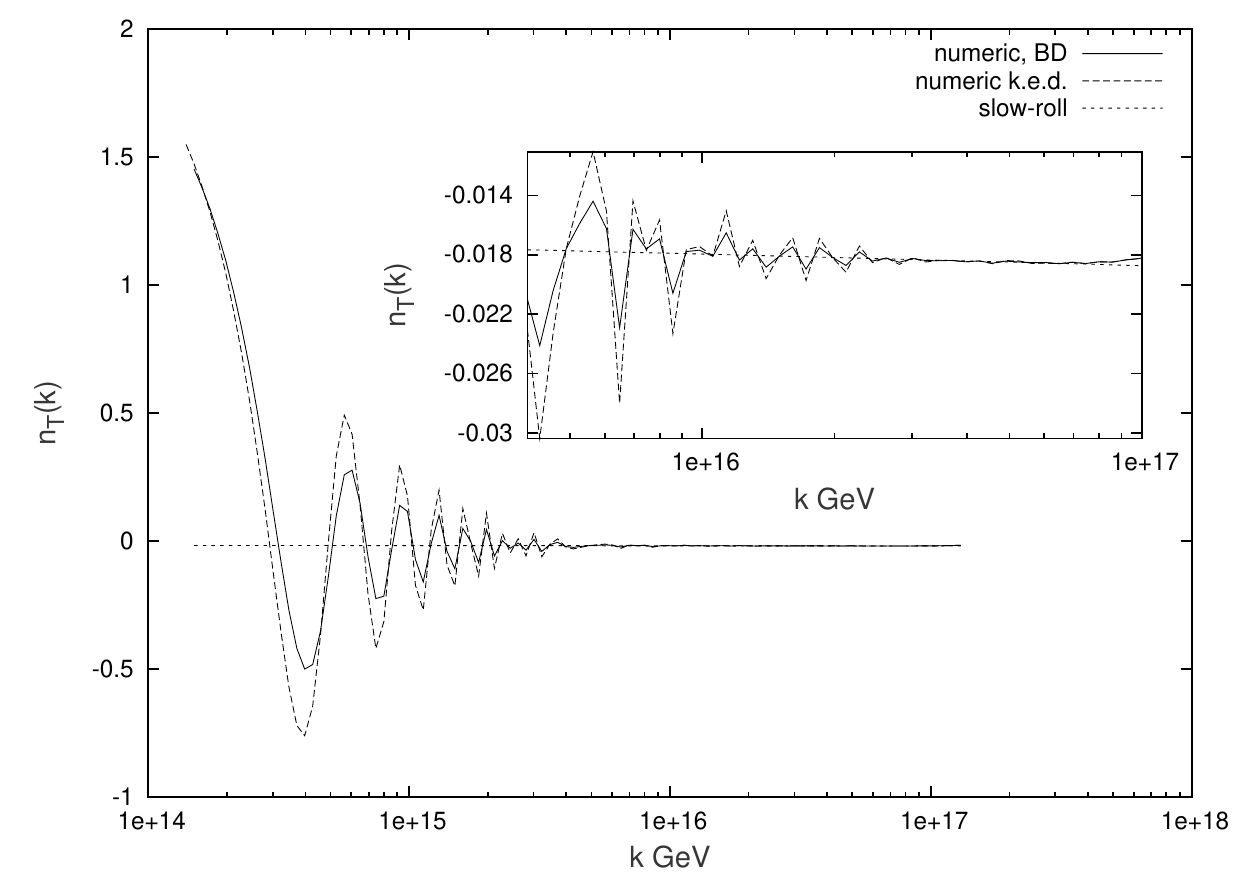}
\includegraphics[height=2.4in,width=3.2in,angle=0]{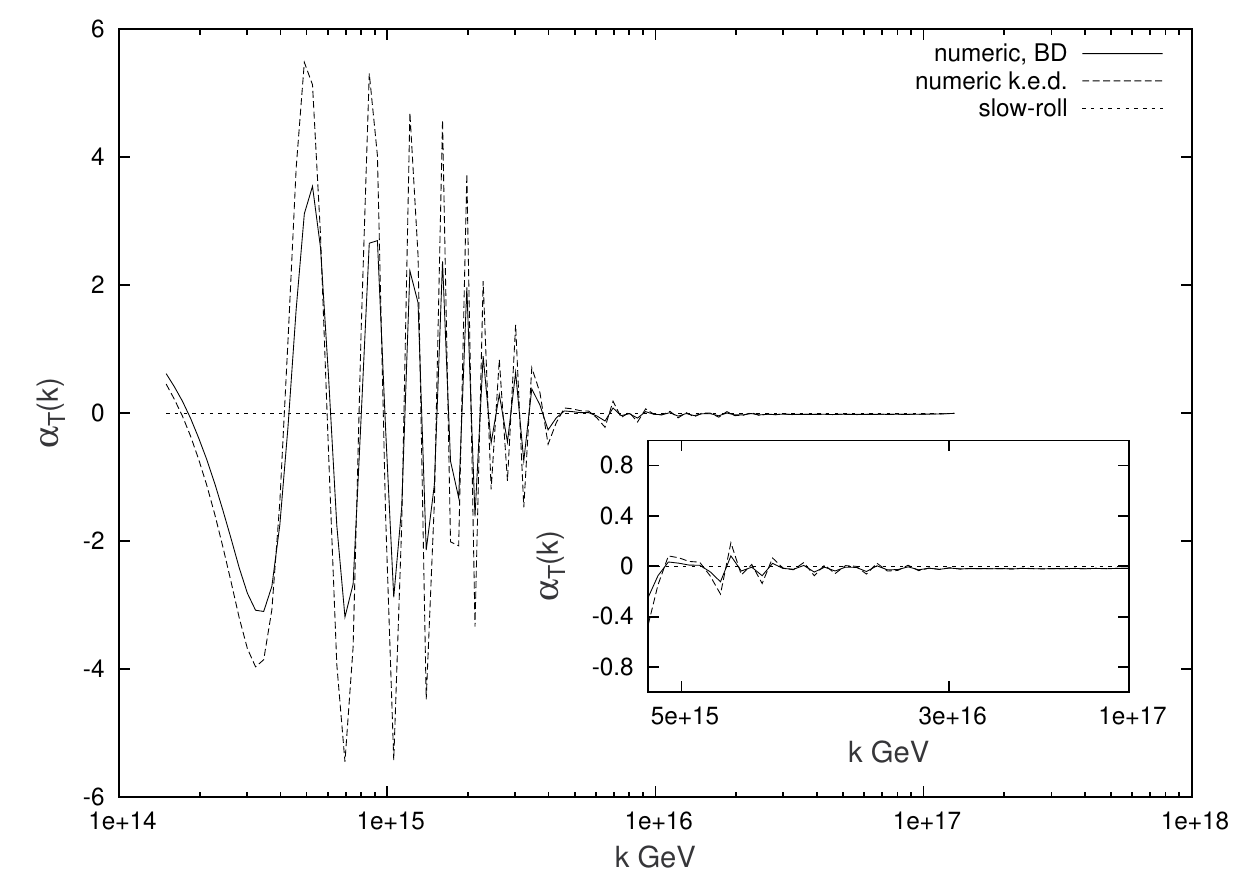}
\caption{Behavior of the spectral index and its running for the tensor modes 
using the mode integration, from the best-fit value 
$\epsilon_{1,i} =2.78$, $\phi_i/M_p = 17.47$,  $6.55\times 10^{-6}$ 
with tensors, cases 1 and 2 in Table~\ref{results}, compared to 
the slow-roll prediction at first order for the same initial condition.}
\label{mode2_t}
\end{centering}
\end{figure*}

\begin{figure}
\begin{centering}
\includegraphics[height=2.4in,width=3.2in,angle=0]{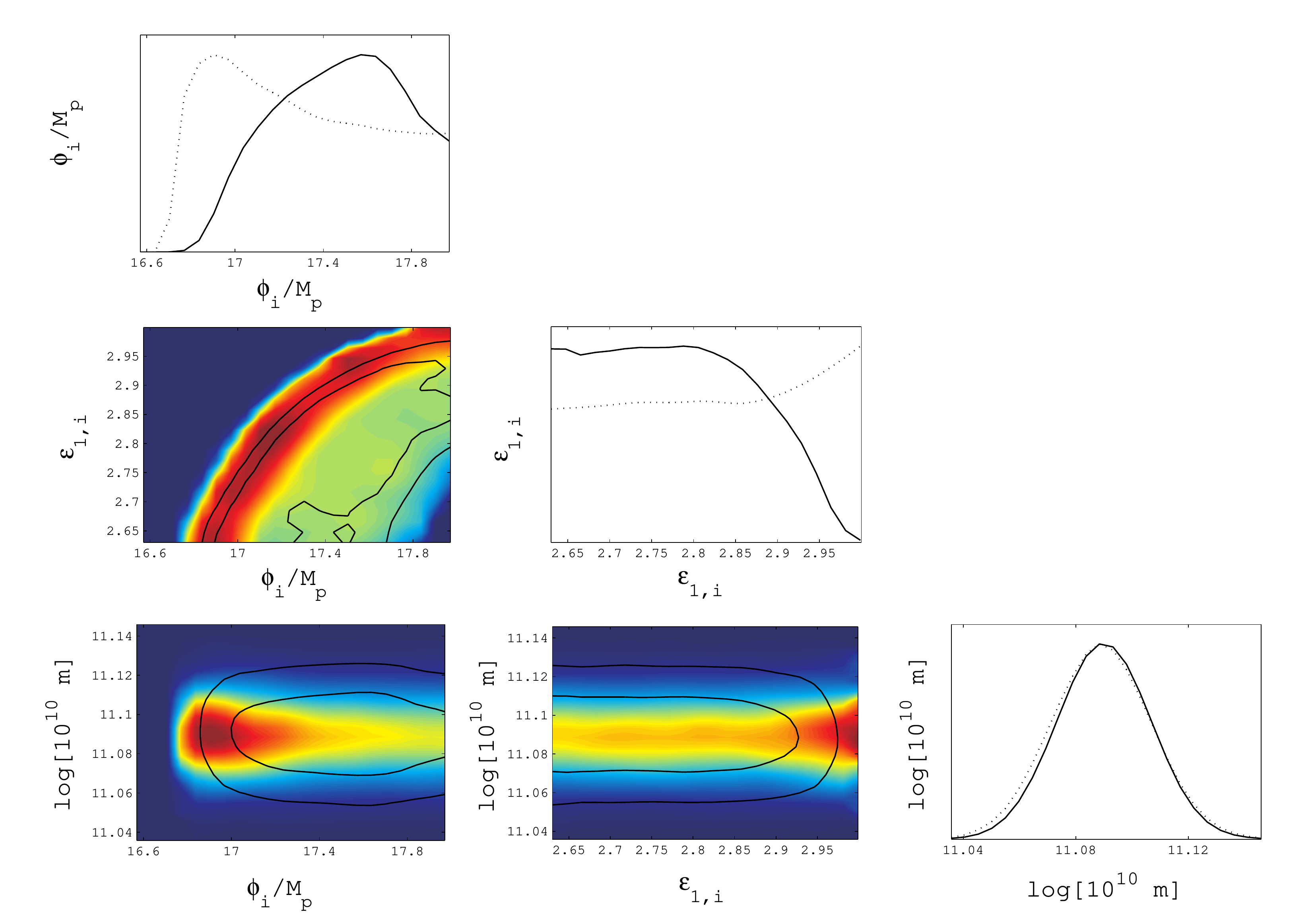}
\caption{(color online). Three dimensional distributions of cosmological 
parameters for cases 1 and 2 in Table~\ref{results2}.}
\label{dist3d}
\end{centering}
\end{figure}

From the mode integration one obtains the values for the amplitudes 
of the spectra for a series of values of the modes $k$. They are therefore 
discrete functions and the previous expressions if applied, do not give as 
expected a result that should join the slow-roll attractor given by 
Equations~(\ref{sr-e}). One therefore needs to evaluate the derivative
of the spectra as a discrete function directly. In this case, the backwards
difference approximation for the first and second derivatives of the 
power spectra are used. For an arbitrary function $P(k_i)$:
\begingroup
\everymath{\scriptstyle}
\small
\begin{eqnarray} 
\label{def}
\frac{dP(k_i)}{dk_i}&\simeq&\frac{P(k_{i+1})-P(k_i)}{k_{i+1}-k_i}, \\ \nonumber 
\frac{d^2P(k_i)}{dk_i^2}&\simeq& \frac{P(k_{i+2})-2P(k_{i+1})+P(k_i)}
{\left(k_i-k_{i-1}\right)^2}
\end{eqnarray}
\endgroup
after  substituting in Equations~(\ref{ir-def}):
\begingroup
\everymath{\scriptstyle}
\small
\begin{eqnarray} 
\label{oe}
n_S = 1+ \frac{k}{P_R}\frac{dP_R}{dk}, \quad
n_T = \frac{k}{P_T}\frac{dP_T}{dk} \qquad \qquad \qquad \qquad\\ \nonumber
\alpha_{S,T} = k\frac{dn_{S,T}}{dk}
=\frac{k^2}{P_{R,T}}\frac{d^2P_{R,T}}{dk^2}
+\frac{k}{P_{R,T}}\frac{dP_{R,T}}{dk} \\ \nonumber
-\left(\frac{k}{P_{R,T}}\frac{dP_{R,T}}{dk}\right)^2. 
\end{eqnarray}
\endgroup
In this case, the spectral indices are evaluated from the second mode
onwards and the runnings from the third onwards. The behavior of the
spectral index and its running for the tensor modes are shown in 
Fig.~\ref{mode2_t}. The distributions of cosmological parameters for case 1
in Table~\ref{results} are presented in Fig.~\ref{dist3d}

\end{document}